\documentclass[sigconf, nonacm, authorversion]{acmart}
\usepackage{algorithm}
\usepackage{algpseudocode}
\usepackage{graphicx}
\usepackage{multirow}
\usepackage{tabularx}
\usepackage{nicematrix}

\AtBeginDocument{%
  }

\begin{document}

\title{HELP: Hierarchical Embeddings-based Log Parsing}

\author{Andy Xu}
\orcid{0009-0008-9461-1545}
\affiliation{%
  \institution{Iudex AI}
  \institution{Harvey Mudd College}
  \city{Claremont}
  \state{California}
  \country{USA}
}
\email{andxu@hmc.edu}

\author{Arno Gau}
\orcid{0009-0005-5392-5511}
\affiliation{%
  \institution{Iudex AI}
  \city{San Fransisco}
  \state{California}
  \country{USA}
}
\email{arno@iudex.ai}

\renewcommand{\shortauthors}{Xu et al.}

\begin{abstract}
  Logs are a first-hand source of information for software maintenance and failure diagnosis. Log parsing, which converts semi-structured log messages into structured templates, is a prerequisite for automated log analysis tasks such as anomaly detection, troubleshooting, and root cause analysis. However, existing log parsers fail in real-world systems for three main reasons. First, traditional heuristics-based parsers require handcrafted features and domain knowledge, which are difficult to generalize at scale. Second, existing large language model-based parsers rely on periodic offline processing, limiting their effectiveness in real-time use cases. Third, existing online parsing algorithms are susceptible to log drift, where slight log changes create false positives that drown out real anomalies. To address these challenges, we propose HELP, a Hierarchical Embeddings-based Log Parser. HELP is the first online semantic-based parser to leverage LLMs for performant and cost-effective log parsing. We achieve this through a novel hierarchical embeddings module, which fine-tunes a text embedding model to cluster logs before parsing, reducing querying costs by multiple orders of magnitude. To combat log drift, we also develop an iterative rebalancing module, which periodically updates existing log groupings. We evaluate HELP extensively on 14 public large-scale datasets, showing that HELP achieves significantly higher F1-weighted grouping and parsing accuracy than current state-of-the-art online log parsers. We also implement HELP into Iudex's production observability platform, confirming HELP's practicality in a production environment. Our results show that HELP is effective and efficient for high-throughput real-world log parsing.
\end{abstract}





\maketitle
\section{INTRODUCTION}
Modern software systems consist of a multitude of interdependent services and components that have become increasingly complex. When these systems experience downtime, this can cause significant revenue loss, especially for large-scale distributed systems \cite{he_survey_2021}, with hourly costs of service outages ranging from hundreds of thousands to millions of U.S. dollars \cite{rapoza_preventing_2014, ponemon_cost_2016}.

Software logs have been widely deployed across various reliability assurance systems \cite{he_survey_2021} as they are often the only data available for a variety of downstream tasks such as anomaly detection \cite{le_log-based_2022, yang_semi-supervised_2021, zhang_robust_2019}, failure troubleshooting \cite{chen_pathidea_2022, xu_largescale_2009}, and root cause analysis \cite{chuah_diagnosing_2010, notaro_logrule_2023}. Therefore, log analysis plays a critical role in maintaining the stability and security of software systems. The first step in log analysis is log parsing, whereby semi-structured log messages are converted into structured log events \cite{khan_guidelines_2022}. Log messages typically consist of 1) \textit{log templates} - fixed text often written by developers that describe the main portion of logged events and 2) \textit{log parameters} - dynamic values of  program variables that carry runtime information from different executions \cite{jiang_lilac_2024}. For example, the log "\texttt{start processing 2 alerts for org org\textunderscore bff943b3ca}" would be parsed into the log template "\texttt{start processing <*> alerts for org <*>}", where the log parameters "\texttt{2}" and "\texttt{org\textunderscore bff943b3ca}" represent the number of alerts and the org id respectively.

A straightforward approach to log parsing would be to match logs with their corresponding source code. However, in practice, source code is often unavailable, particularly for commercial software and third-party libraries. An alternative is to perform manual log parsing, but due to the large volume of logs generated—up to hundreds of millions of logs per hour—this is error-prone and impractical \cite{wang_spine_2022, he_survey_2021}. Therefore, numerous automated log parsers have emerged in recent years. These parsers can be categorized into traditional heuristics-based parsers \cite{he_drain_2017, du_spell_2016, dai_logram_2022} and deep learning semantic-based parsers \cite{liu_uniparser_2022, jiang_lilac_2024, xu_divlog_2024, le_log_2023}. 

Although many of the aforementioned log parsing methods have shown strong performance in previous benchmarks, recent  studies \cite{khan_guidelines_2022, jiang_large-scale_2023, petrescu_log_2023}  have shown that existing log parsers fail in high throughput real world systems, particularly those with a large and diverse set of log templates. This is because heuristics-based parsers rely on hand-crafted rules and specific domain knowledge \cite{he_drain_2017, du_spell_2016, dai_logram_2022} while semantic-based log parsers often require labeled data which prevents the algorithm from working effectively when applied to new data sources \cite{le_log_2023, liu_uniparser_2022}. 

In addition, existing state-of-the-art log parsers are offline, rather than online algorithms \cite{jiang_lilac_2024, xu_divlog_2024, he_survey_2021}. Offline log parsers require all log messages a priori and batch parse log messages periodically. In contrast, online log parsers parse log messages in a streaming manner. This makes online log parsing preferable for downstream applications, as many of these applications benefit from real-time parsing that can flag and resolve errors quickly to avoid system downtime. However, this also makes online parsing more difficult due to their incomplete access to log context \cite{shima_lenma_2016}.

Therefore, we propose HELP, a \textbf{H}ierarchical \textbf{E}mbeddings-based \textbf{L}og \textbf{P}arser. HELP is the first framework to utilize semantic embeddings and vector databases for log parsing. To create domain-specific embeddings, we fine-tuned a universal embedding model on over 300,000 logs. By separating log clustering and log parsing, HELP makes online LLM-based parsing performant and cost-effective. 

HELP consists of three main components. The first component is the online hierarchical embedding module, which takes in raw logs and word frequency statistics, generates vector embeddings for logs from their semantic and structural meaning, and clusters logs into their corresponding patterns. The second component is the context-aware parsing module, which leverages few-shot and Chain of Thought (CoT) prompting in Large Language Models (LLMs) to extract log templates. The final component is the iterative rebalancing module, which periodically merges log clusters and reorients vector embeddings to improve grouping accuracy and prevent template drift. By utilizing cosine similarity, language models, and word frequency statistics, HELP combines the advantages of clustering, semantic, and heuristics-based parsers.

We evaluated the performance of HELP against other state-of-the-art online and offline log parsers on 14 large-scale log parsing datasets in Loghub-2.0 \cite{jiang_large-scale_2023} from the LogPAI team\cite{zhu_tools_2019}. Against the state-of-the-art online parsers, Drain and UniParser, HELP achieves significantly higher accuracy across all metrics with 62.9\% and 79.2\% higher F1 weighted grouping accuracy (FGA), and 169.0\% and 189.6\%  higher weighted template accuracy (FTA) respectively. Against the state-of-the-art parser, LILAC, HELP achieves comparable performance despite being an online algorithm. Moreover, HELP is highly parallelizable due to it's rebalancing module, with the ability to batch ingest thousands of logs with little to no performance degradation. Due to it's efficiency and accuracy, we successfully integrated HELP into Iudex's production observability platform. Our evaluation results confirm that HELP can process logs in real-time, achieving a 1.5 second P95 end-to-end latency from log emission to pattern assignment.

To summarize, our main contributions are as follows: 
\begin{itemize}
\item We propose HELP, the first semantic embeddings-based log parser for performant and cost-effective online log clustering and parsing. 
\item We evaluated HELP on 14 public log datasets. Our experimental results show that HELP outperforms all other state-of-the-art log parsers in log grouping and parsing, and can easily be converted into a parallelizable batch processing framework with little performance degradation.
\item We confirmed the practicality and efficiency of HELP by deploying HELP in a production environment. HELP achieves real-time pattern assignment and avoids template drift via its periodic rebalancing module.

\end{itemize}

\section{BACKGROUND AND MOTIVATION}
\subsection{Log Parsing}
Log parsing is the initial and most critical step in log analysis \cite{he_survey_2021, zhu_tools_2019}, whereby parsers convert semi-structured log messages into structured data. This is done by extracting the constant (log templates) and dynamic (log parameters) parts from log messages. Although some studies have extracted log templates from logging statements in source code \cite{pecchia_industry_2015, schipper_tracing_2019}, this is often impractical when source code is inaccessible, such as in commercial software and third-party libraries \cite{zhu_tools_2019}. Consequently, many automated log parsers have emerged in recent years. These data-driven log parsing approaches can be divided into four main groups:
\begin{enumerate}
    \item \textit{Frequent pattern mining}. Offline log parsers like SLCT \cite{vaarandi_SLCT_2003}, LFA \cite{nagappan_LFA_2010}, and Logram \cite{dai_logram_2022} find frequent patterns that emerge across the entire dataset, leveraging token position or \textit{n}-gram information to extract log templates. 
    \item \textit{Clustering}. Offline log parsers like LogCluster \cite{vaarandi_logcluster_2015} and LogMine \cite{hamooni_logmine_2016}, and online parsers like LenMa \cite{shima_lenma_2016} group similar logs together via their token frequencies or similarity. They consider logs clustered together as belonging to the same template.
    \item \textit{Heuristics}. Offline log parsers like AEL \cite{jiang_ael_2008}, and online parsers like Spell \cite{du_spell_2016} and Drain \cite{he_drain_2017} employ heuristics based on assumptions about log structure and token counts to extract templates efficiently.
    \item \textit{Semantic}. Recent offline log parsers like LogPPT \cite{le_log_2023} and LILAC \cite{jiang_lilac_2024} utilize the broad pre-trained knowledge of Large Language Models (LLMs), sometimes combined with additional fine-tuning, to effectively parse log templates. 
\end{enumerate}
\subsection{Large Language Models}
Large Language Models (LLMs) have shown remarkable performance for various natural language processing tasks. These models adopt the Transformer architecture \cite{vaswani_attention_2017} and are pre-trained on a vast quantity of text corpora using self-supervised objectives \cite{brown_language_2020}. Recent studies have highlighted that LLM-based parsers \cite{xu_divlog_2024, le_log_2023, jiang_lilac_2024} can significantly outperform other log parsers \cite{khan_guidelines_2022, jiang_large-scale_2023, petrescu_log_2023}. Most of these methods (e.g. DivLog \cite{xu_divlog_2024} and LILAC \cite{jiang_lilac_2024}) utilize the in-context learning (ICL) paradigm of language models, wherein LLMs can be applied to downstream tasks without the resource intensive task of fine-tuning model parameters \cite{dong_survey_2022, liu_pre-train_2023}. They first group logs based on log characteristics like cosine similarity \cite{xu_divlog_2024} or top-K frequent tokens \cite{jiang_lilac_2024}, then sample candidates to maximize similarity or diversity.

\begin{figure}[t!]
  \centering
  \includegraphics[width=\linewidth]{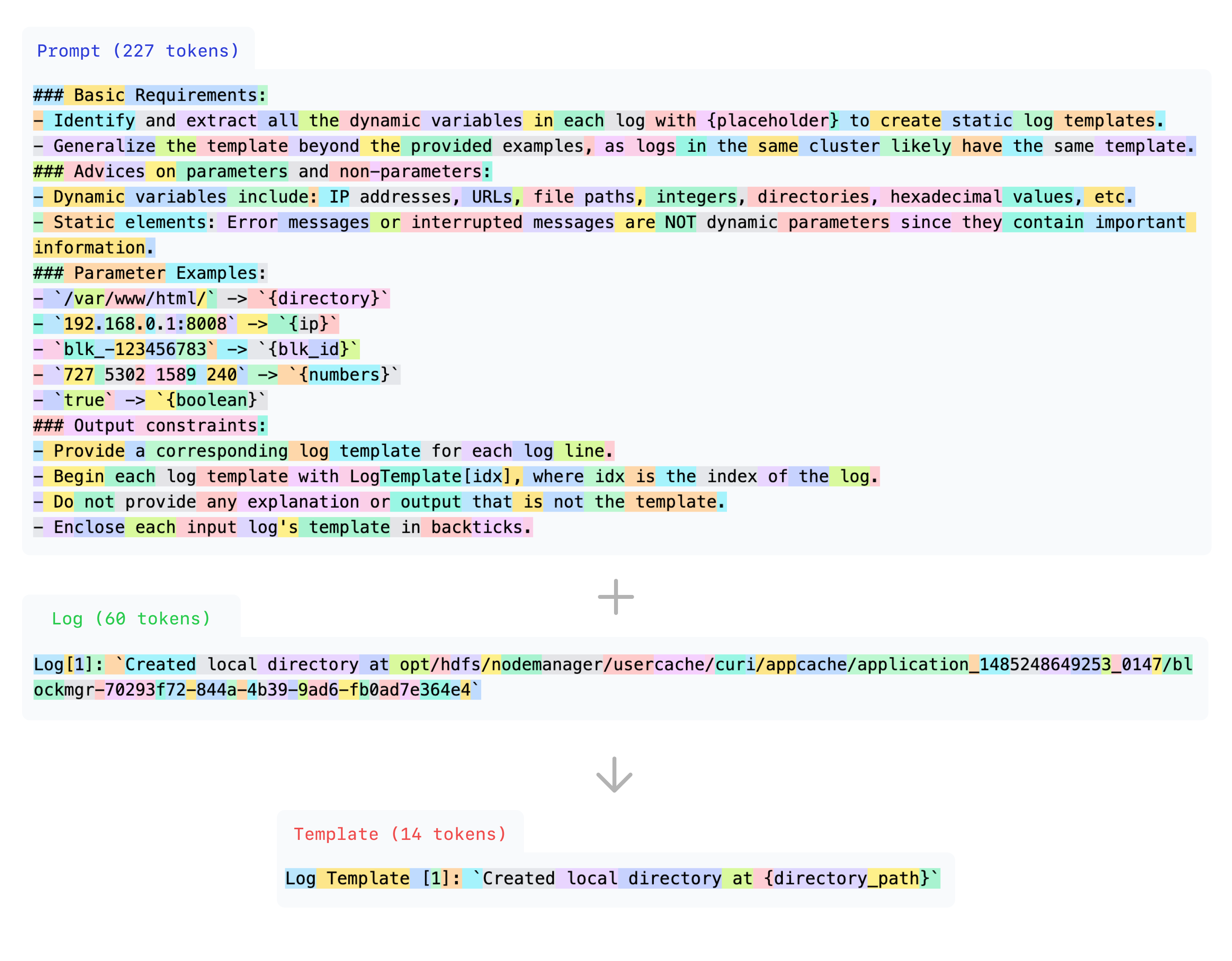}
  \caption{Example zero shot LLM query with tokenizer visualization.}
  \Description{Visualization of zero shot LLM query with the prompt, log and output template and their corresponding tokens }
  \label{fig:token}
\end{figure}

However, despite promising results, all LLM-based parsers fail to support online parsing. Since LLMs have billions of weights and require extensive GPU compute for inference, they are often deployed on high-performance servers and are queried through APIs. This makes sequential querying of LLMs \cite{xu_divlog_2024, le_gpt_2023} impractical in production deployment for its latency and cost \cite{jiang_large-scale_2023, mudgal_assessment_2023}. For example, as shown in ~\autoref{fig:token}, a zero shot query similar to the one used by LILAC would required 227 tokens for the prompt template and 60 tokens for the log itself per OpenAI's tokenizer \cite{openai_tokenizer}. This generates an output for the log template that requires 14 tokens \cite{openai_tokenizer}. This means that for every million logs, the cost of querying OpenAI's cheapest model GPT-4o mini would be $(227+60)*1,000,000*(0.15/1,000,000) + (14)*1,000,000*(0.6/1,000,000)= \$51.45$ per 1M logs, assuming the price of GPT-4o mini is \$0.15 per 1M input tokens and \$0.60 per 1M output tokens \cite{openai_pricing}. Given that systems can generate up to hundreds of millions of logs per hour \cite{wang_spine_2022, he_survey_2021}, this can cost hundreds of thousands of dollars a day, making LLM-based online parsing infeasible in practice.

\subsection{Embedding Models}
Although sequential querying of LLMs remains expensive and infeasible at scale, embedding models represent an alternative for log clustering. Embedding models enable a similar high-level semantic understanding of logs while requiring significantly fewer computational resources, reducing latency and cost. Embedding models can be categorized into two groups: word and text embeddings. Word embeddings represent each individual word as its own vector, while text embeddings compress variable length text sequences into a single vector representation. Although previous methods have applied word embeddings for log clustering and anomaly detection \cite{vaarandi_logcluster_2015, meng_2019_loganomaly, meng_2020_semanticaware, xiao_2024_stronger}, text embeddings are advantageous due to their contextual understanding across the entirety of a log and their invariance to log length. One such text embedding model is OpenAI's text-embedding-3-small, which is initialized from pre-trained GPT models and refined via unsupervised contrastive training for semantic similarity \cite{neelakantan_2022_text}. Compared to the cheapest completion model, OpenAI's text embedding model is an order of magnitude cheaper by token \cite{openai_pricing}. Since no prompt or output is required, much fewer tokens are required, further reducing the cost of embedding models. This makes embedding models uniquely suited to support online log clustering, enabling downstream cluster-based log parsing.

\section{METHODOLOGY}
\begin{figure}[t!]
  \centering
  \includegraphics[width=0.8\linewidth, trim=10 10 10 10, clip]{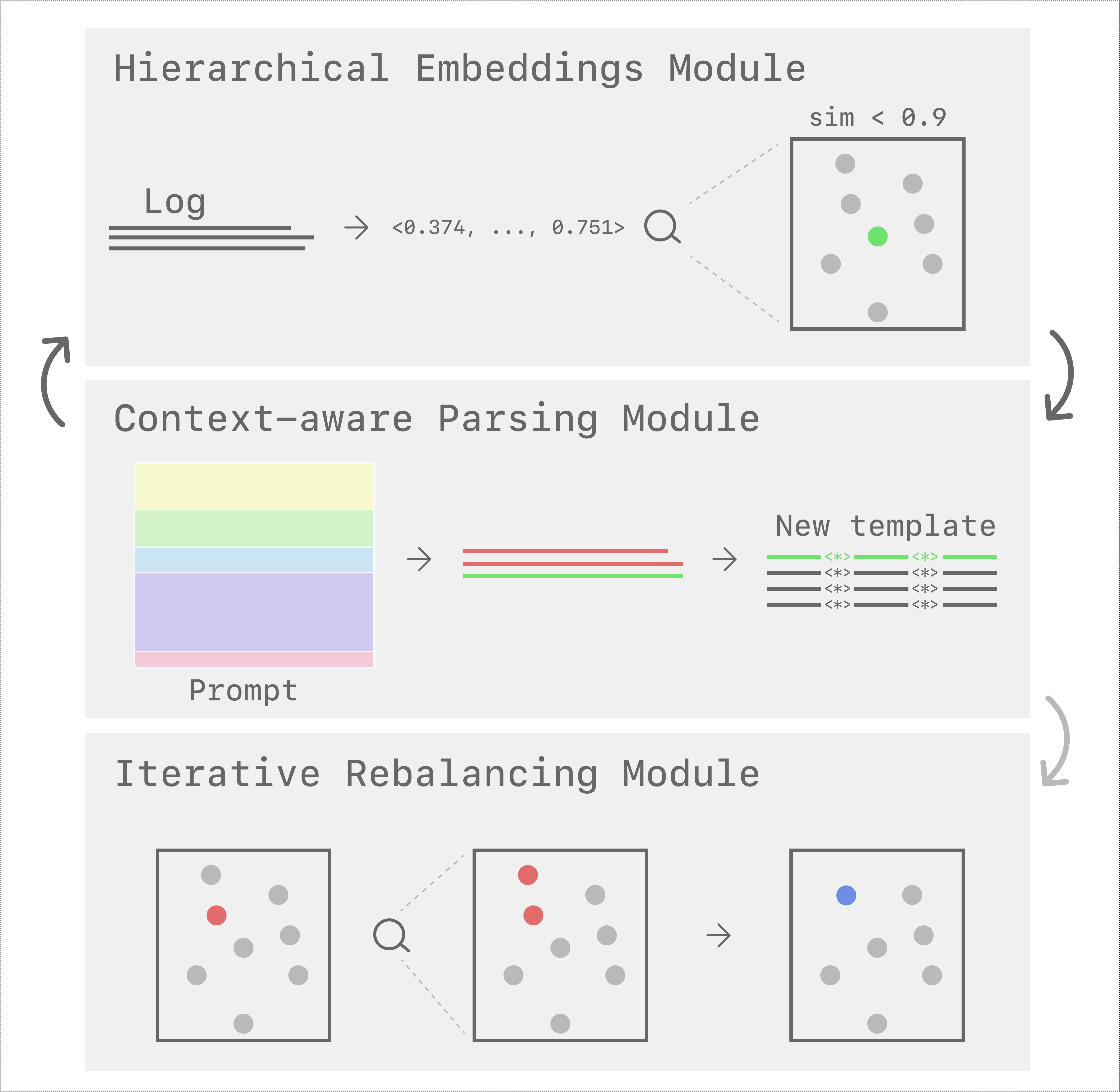}
  \caption{Overview of HELP components when inserting a new log template}
  \Description{Visualization of the hierarchical embeddings module, context-aware parsing module, and iterative rebalancing module }
  \label{fig:overview}
\end{figure}
\subsection{Overview}
\label{sec:overview}
In this section, we introduce HELP, a Hierarchical Embeddings-based Log Parser. HELP consists of three main components: the online hierarchical embedding module, the context-aware parsing module, and the iterative rebalancing module. To specialize HELP for log data, we train a neural network encoder on over 300,000 logs which we stack on top of the universal OpenAI embedding model. We further utilize the in-context learning paradigm of the LLM by providing few-shot examples of parsed logs coupled with chain of thought reasoning to explain parsing rationale. To enable online parsing, HELP introduces a novel efficient and cost-effective embeddings-based clustering pipeline. HELP is built on the premise that there exist many orders of magnitude fewer log templates than logs in real-world systems \cite{jiang_large-scale_2023, wang_spine_2022}. For example, the datasets in Loghub-2.0 contain over 50 million logs, but the total number of templates is less than 3,500 \cite{jiang_large-scale_2023}. Therefore, by first clustering logs and querying an LLM once per cluster, HELP significantly increases parsing efficiency, enabling the first LLM-based online parsing method. Via its iterative rebalancing module, HELP can reflect on its existing log groupings, combining similar clusters and avoiding template drift.

~\autoref{fig:overview} is an overview of HELP's online ingestion pipeline. HELP first takes in raw logs and word frequency statistics, and generates a custom vector embedding. Then, HELP performs an approximate nearest neighbor (ANN) search of the vector database containing existing log templates. If the cosine similarity is below $T_c$, HELP will insert the new vector into the vector database. Otherwise, HELP will merge the two vectors together, and the existing log template of the group will be assigned to the vector. In addition, the parsing module will be called, which will parse and create a log template based on the log in the new cluster. For the offline rebalancing process invoked every $N$ logs, the rebalancing module will sequentially search for the closest neighbor for every vector and merge them if their cosine similarity is above $T_c$.

\subsection{Hierarchical Embeddings Module}
Directly querying an LLM for online log parsing is expensive and inefficient due to the large volume of log data. Therefore, HELP first embeds logs into vectors, and then clusters vectors based on their cosine similarity.

\subsubsection{Embedding}
While the semantic meaning of logs is crucial for clustering, relying solely on semantic embeddings can create overly specific groups (e.g. groups based on users or filepaths) that do not generalize well to log templates. To refine the embeddings for log template clustering, we make two key modifications to OpenAI's universal embeddings:
\begin{algorithm}[t]
\caption{Hierarchical Embeddings Module}\label{alg:hem}
\begin{algorithmic}[1]
\Require Set of logs $L = \{x_i\}_{i=1}^n$, similarity threshold $T_c$, OpenAI embedding $\epsilon(\cdot)$, nn encoder $\phi(\cdot)$
\State Initialize set of existing vectors $\mathcal{V} = \emptyset$
\For{$x_i \in L$}
    \State $v'_i = \epsilon(x_i)$
    \State $v_i = \phi(v'_i)$
    \State $v_i = \frac{v_i}{\|v_i\|}$ 
    \State $(v_j, c) = \text{simSearch}(v_i, \mathcal{V})$
    \If{$c < T_c$}
        \State $\mathcal{V} = \mathcal{V} \cup \{v_i\}$
    \Else
        \State $w_j = $ getWeight$(v_j)$
        \State $v_j = v_j + \frac{v_i - v_j}{w_j+1}$
        \State $v_j = \frac{v_j}{\|v_j\|}$ 
        \State Update $v_j$ in $\mathcal{V}$ 
    \EndIf
\EndFor
\end{algorithmic}
\end{algorithm}

\textbf{1) Word count:} Previous work have shown the correlation between the word count of a log message and its log template \cite{shima_lenma_2016}. However, LLMs and embedding models struggle to accurately determine word count due to their byte-pair encoding, where individual tokens can consist of parts of words or multiple words \cite{neelakantan_2022_text}. Empirically, LogPAI \cite{jiang_large-scale_2023} showed that semantic-based log parsers are superior at log parsing, but are weaker at grouping because of their lack of global information. Therefore, HELP embeds the word count of each log in addition to its log content. This improves clustering accuracy as the word count of the current log is compared with previous ingested logs. Thus, the ingestion process considers both macro-level semantic and statistical features of log messages. 

\textbf{2) Custom Embedding:} To fine-tune OpenAI's embedding model for log template clustering, we stack a two layer linear neural network encoder on top of the OpenAI embedding output. This allows HELP to fine-tune OpenAI's embedding model without needing access to the weights of the black box model, making the process cheaper, easier, and less data-intensive. The neural network is trained via 14 fold cross validation, where the model is trained and validated on 13 of the datasets, and tested on the last dataset. The two layer linear neural network architecture was chosen due to its simplicity, which preserves the rich semantic information present in the universal OpenAI embeddings and reduces the risk of overfitting. Each dataset consists of OpenAI embeddings log pairs, labeled as similar (1) or dissimilar (0) based on their log patterns. Each embedding is passed through the same neural network, and the model minimizes the Mean Squared Error (MSE) loss between the predicted cosine similarity of the neural network encoder embeddings and the ground truth labels. The MSE loss function is defined as follows:
\begin{equation}
\text{MSE} = \frac{1}{N} \sum_{i=1}^{N} (y_i - \hat{y}_i)^2
\end{equation}
where \( y_i \) are the target values, \( \hat{y}_i \) are the predicted values, and \( N \) is the number of pairs. The complete list of hyperparameters used for training and testing are described in ~\autoref{sec:implementation}.

\subsubsection{Vector Update}
As mentioned in ~\autoref{sec:overview}, there exist orders of magnitude fewer log templates than logs. Therefore, in order to reduce storage and computational costs, HELP stores one vector per cluster. This means, however, that traditional clustering algorithms like K-Means \cite{ikotun_2023_kmeans} and DBSCAN \cite{ester_1996_densitybased} cannot be applied. Instead, HELP employs an incremental vector update that approximates the centroid of each cluster via a moving average. 

The detailed algorithm is shown in ~\autoref{alg:hem}. For each log $x_i$ in the set of logs $L$, HELP first encodes the log using the OpenAI embedding $\epsilon(\cdot)$ to obtain an intermediate vector $v'_i$. This vector is passed through the neural network encoder $\phi(\cdot)$ and normalized, resulting in the final vector $v_i$. HELP then performs a similarity search $\text{simSearch}(v_i, \mathcal{V})$ across the Hierarchical Navigable Small Worlds (HNSW) \cite{malkov_hnsw_2018} graph $\mathcal{V}$, comparing $v_i$ against the existing vectors in $\mathcal{V}$ and yielding the closest match $v_j$ and corresponding similarity score $c$. Note that the use of HNSW allows for $O(log(N))$ search \cite{malkov_hnsw_2018}, which scales better for large datasets than a naive k-Nearest Neighbor (k-NN) $O(N)$ search. If the similarity score $c$ is below the threshold $T_c$, the vector $v_i$ is added to the set $\mathcal{V}$. Otherwise, the existing vector $v_j$ is updated in $\mathcal{V}$ such that $v_j = v_j + \frac{v_i - v_j}{n+1}$, where $n= $ getWeight$(v_j)$ is the number of logs in the cluster corresponding to $v_j$.

\subsection{Context-Aware Parsing Module}

\begin{figure}[t!]
  \centering
  \includegraphics[width=\linewidth]{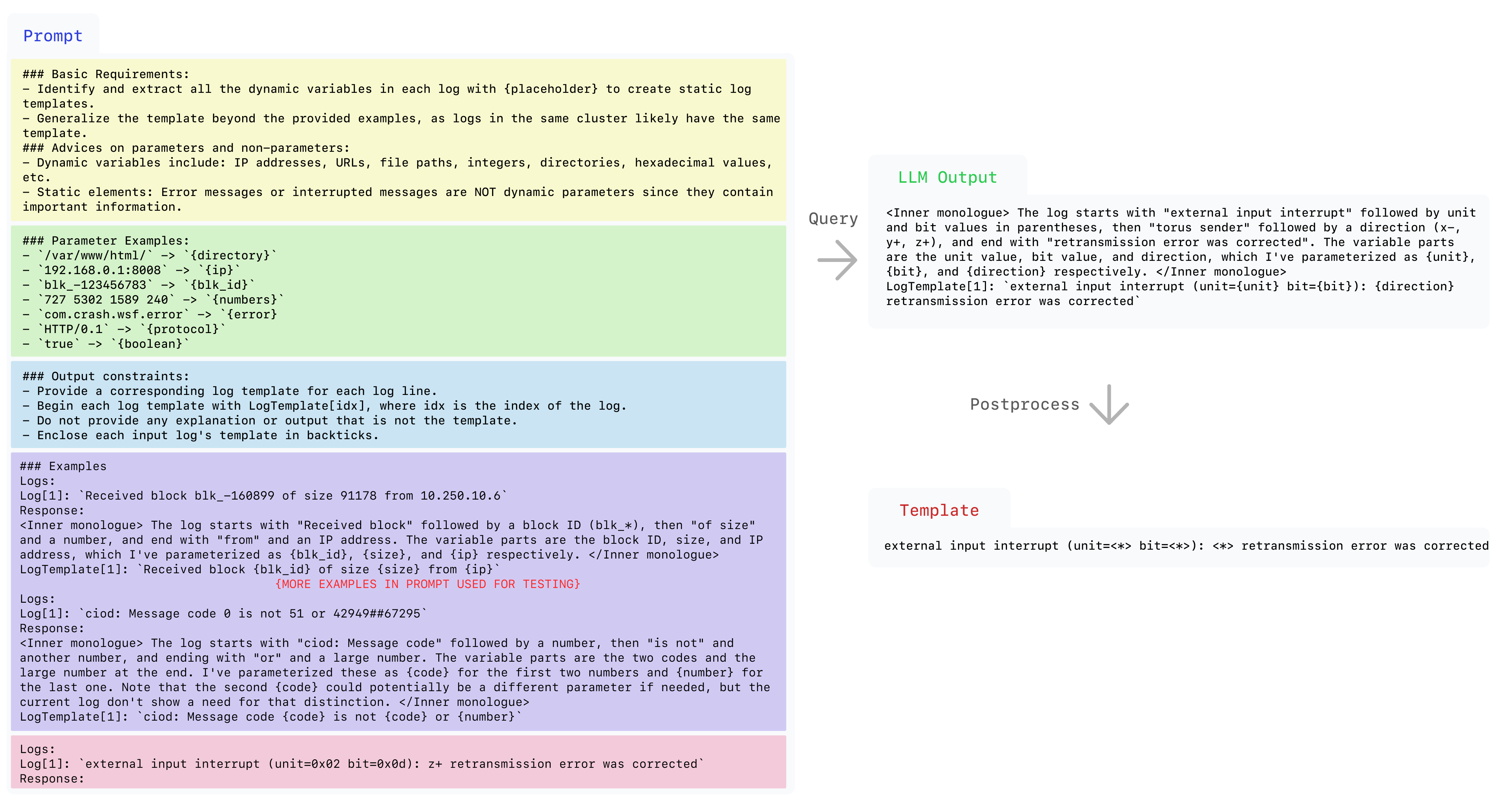}
  \caption{Context-aware parsing module template generation example.}
  \Description{Example of parsing module full prompt, the generated LLM output, and the final parsed template.}
  \label{fig:prompt}
\end{figure}
When HELP creates a new log cluster, the context-aware parsing module parses the log via the state-of-the-art LLM, Claude-3.5-Sonnet \cite{anthropic_2024_claude}, where the resultant log template will be assigned to all subsequent logs in the cluster. The parser leverages the strong textual and semantic comprehension of LLMs, which have shown remarkable zero-shot and few-shot performance across many natural language processing tasks. Thus, although we do not provide log samples from the same log source in context, we believe that by eliciting strong reasoning capabilities and providing in-domain knowledge, the LLM can still effectively parse log templates.

\subsubsection{Prompt Design}
Following previous work \cite{huang_2024_ulog}, we design the prompt format to query the LLM and generate the log template for individual log messages. Our prompt consists of five parts: task instructions, parameter examples, output constraints, parsing demonstrations, and the queried log. ~\autoref{fig:prompt} shows an example.
\begin{enumerate}
  \item \textbf{Task Instructions.} To provide the LLM with task-specific log information, we utilize a special system prompt to give an overview of log parsing and align the model for log analysis. We also provide instructions that include basic requirements, which specify the task to extract dynamic variables from logs, and advice on parameters, which give descriptors for common dynamic and static elements.
  \item \textbf{Parameter Examples.} To provide domain knowledge to the model, we provide specific examples of common dynamic parameters such as an ip "\texttt{192.168.0.1:8008}" or boolean "\texttt{true}". The use of domain specific placeholders rather than universal \texttt{\{variables\}} allows the LLM to better contextualize and recognize dynamic values.
  \item \textbf{Output Constraints} To ensure the model creates a consistent, parsible output, we specify the desired output, where each template is prepended by "LogTemplate[idx]" and delimited by backticks.
  \item \textbf{Parsing Demonstrations.} Few-shot demonstrations of log parsing have been shown to greatly improve performance on downstream tasks \cite{zhao_2021_calibrate}. In addition, Chain of Thought has been shown to significantly improve the ability of LLMs to perform complex reasoning \cite{wei_2022_chainofthought}. Therefore, HELP provides four examples of logs, their reasoning, and their corresponding log templates. To keep the reasoning brief and self-contained, reasoning is denoted by the pair of tags \texttt{<Inner Monologue>} and \texttt{</Inner Monologue>}.
\item \textbf{Queried Log.} Finally, the log message, prepended by "Log[idx]" is fed into the model.
\end{enumerate} 
\subsubsection{Template Extraction}
After receiving a response from the LLM, we use regex post-processing to extract valid templates designated by backticked strings with a "LogTemplate" prefix similar to ULog \cite{huang_2024_ulog}. Then, we replace all bracketed parameters by the placeholder \texttt{<*>}, and merge common placeholders (e.g. \texttt{<*><*> $\rightarrow$ <*>} following previous work \cite{jiang_lilac_2024, xu_divlog_2024, le_log_2023, liu_uniparser_2022}.

\subsection{Iterative Rebalancing Module}
Because the hierarchical embeddings module performs a vector update which reorients centroids over time, vectors from different clusters can drift closer together, indicating that they should be merged into a single cluster. This is an inherent disadvantage to online clustering methods, where not all information is available at inference. To address this, HELP incorporates an Iterative Rebalancing Module, detailed in ~\autoref{alg:irm} and depicted in ~\autoref{fig:overview}, to periodically reflect and merge existing clusters. 

For each vector $v_i$ in the set of vectors $\mathcal{V}$, the algorithm performs a similarity search $\text{simsearch}(v_i, \mathcal{V})$ to find the closest match $v_j$ and the corresponding similarity score $c$. If the similarity score $c$ is greater than or equal to the threshold $T_c$, the algorithm merges the two vectors. The number of logs in each cluster $w_i$ and $w_j$ of the vectors $v_i$ and $v_j$ are retrieved using the function $\text{getWeight}(\cdot)$. The vectors $v_i$ and $v_j$ are then removed from the set $\mathcal{V}$, and a new vector $v$ is computed as the weighted average of $v_i$ and $v_j$, given by $v = \frac{w_i \cdot v_i + w_j \cdot v_j}{w_i + w_j}$ and normalized. The updated vector $v$ is then added back into the set $\mathcal{V}$. If the similarity score $c$ is less than the threshold $T_c$, the algorithm increments the index $i$ and continues to the next vector.

Since the index $i$ is only incremented if $c < T_c$, this allows for a single vector to be successively merged with multiple other vectors. This process repeats until all vectors in $\mathcal{V}$ have been processed. 

\begin{algorithm}[t]
\caption{Iterative Rebalancing Module}\label{alg:irm}
\begin{algorithmic}[1]
\Require Set of vectors $\mathcal{V} = \{v_i\}_{i=1}^m$, similarity threshold $T_c$
\State Initialize index $i = 0$
\While{$i < |\mathcal{V}|$}
    \State $(v_j, c) = \text{simsearch}(v_i, \mathcal{V})$
    \If{$\text{cosSim} \geq T_c$}
        \State $w_i = \text{getWeight}(v_i)$
        \State $w_j = \text{getWeight}(v_j)$
        \State $\mathcal{V} = \mathcal{V} \setminus \{v_i, v_j\}$
        \State $v = \frac{w_i \cdot v_i + w_j \cdot v_j}{w_i + w_j}$
        \State $v = \frac{v}{\|v\|}$
        \State $\mathcal{V} = \mathcal{V} \cup \{v\}$
    \Else
        \State $i = i + 1$
    \EndIf
\EndWhile
\State \Return $\mathcal{V}$
\end{algorithmic}
\end{algorithm}

\subsection{Parallelization}
In previous sections, we have discussed how HELP can be deployed in a sequential fashion. However, in practice, multiple services can emit logs concurrently, requiring multiple logs to be processed at once. Therefore, we also deploy HELP in an online, batched fashion, whereby the hierarchical embeddings module can embed, search, and upload vectors into the vector database in parallel. 

A key edge-case with the proposed parallelization arises when multiple logs with the same unseen log pattern are processed simultaneously. Since logs are processed in parallel and cannot "see" each other in the similarity search, each log will independently return a similarity score less than the threshold. This causes duplicates of the same pattern to be inserted. The iterative rebalancing module resolves this issue by combining all logs that should belong to the same cluster. To reduce querying costs when deploying HELP in parallel, the context-aware parsing module processes new patterns after the iterative rebalancing module has merged existing vectors, rather than at insertion into the vector database.
\section{EXPERIMENTAL DESIGN}
\subsection{Research Questions}
We evaluate our approach by answering the following research questions (RQs):
\begin{itemize}
    \item \textbf{RQ1: } How effective is HELP?
    \item \textbf{RQ2: } How does each component contribute to HELP?
    \item \textbf{RQ3: } How practical is HELP in a production environment?
\end{itemize}
\subsection{Datasets and baselines}
We evaluate HELP on Loghub-2.0 \cite{zhu_2023_loghub, jiang_large-scale_2023}, a collection of large-scale benchmark datasets for log parsing from the LogPAI team \cite{zhu_tools_2019}. Loghub-2.0 contains ground-truth templates for 14 diverse log datasets, covering distributed systems, supercomputer systems, and server-side applications. On average, each dataset in Loghub-2.0 contains 3.6 million log messages, and there are approximately 3,500 unique log templates in total.

We compare HELP against six open-sourced state-of-the-art log parsers, including three online and three offline parsers. For the online parsers, we implement two syntax-based methods, LenMa \cite{shima_lenma_2016}, and Drain \cite{he_drain_2017}, and one semantic-based method, UniParser \cite{liu_uniparser_2022}. LenMa and Drain were chosen for their superior performance compared to other syntax-based online parsers \cite{jiang_large-scale_2023, khan_guidelines_2022, zhu_tools_2019}. UniParser is the state-of-the-art semantic-based online parser and trains a bidirectional long-short term memory (BiLSTM) model on labeled log data for log parsing. For the offline parsers, we select one syntax-based method AEL \cite{jiang_ael_2008} and two semantic-based methods LogPPT \cite{le_log_2023} and LILAC \cite{jiang_lilac_2024}. LogPPT uses labeled log data for prompt-based fine-tuning of RoBERTa while LILAC utilizes the ICL paradigm of LLMs to parse logs by sampling from $k$ similar logs. 

\subsection{Metrics}
\label{sec:met}
Following previous studies \cite{jiang_large-scale_2023, khan_guidelines_2022, zhu_tools_2019}, we evaluate all methods using the following four metrics:
\begin{itemize}
    \item \textit{Grouping Accuracy (GA)} is a log-level metric that is computed as the ratio of correctly grouped log messages over all log messages. A log message is considered grouped correctly if and only if its predicted template have the same group of log messages as the ground truth.
    \item \textit{F1 score of Grouping Accuracy (FGA)} is a template-level metric that measures the ratio of correctly grouped templates. Let $N_g$ be the correct number of templates in the ground truth dataset and $N_p$ be the number of templates generated by the log parser. Suppose $N_c$ is the number of templates correctly parsed by the log parser. Then, Precision of Grouping Accuracy $PGA = \frac{N_c}{N_p}$ and Recall of Grouping Accuracy $RGA = \frac{N_c}{N_g}$. FGA is their harmonic mean, where $FGA = \frac{2\times PGA\times RGA}{PGA+RGA}$
    \item \textit{Parsing Accuracy (PA)} is a log-level metric that measures the correctness of extracted templates. PA is computed as the proportion of correctly parsed log messages to the total number of log messages. A log message is correctly parsed if and only if all tokens are identical with the ground truth.
    \item \textit{F1 score of Template Accuracy (FTA)} is a template-level metric that computes the harmonic mean of precision and recall of template accuracy. A template is considered correct if and only if log messages of the parsed template have the same group as the ground-truth template logs and all tokens in the template are the same as the ground-truth.
\end{itemize}

\subsection{Environment and Implementation}
\label{sec:implementation}
We conduct all experiments on an M3 Macbook Air with 24 GB of RAM. We use the text-embedding-3-small model for embeddings due to its cheap inference costs and strong performance \cite{openai_pricing}. We use Anthropic's Claude-3.5-Sonnet model from July 2024 \cite{anthropic_2024_claude} because of its superior performance compared to OpenAI's GPT models, and set its temperature to 0 so the same output would be generated per query to ensure reproducibility and consistency. 

For the neural network encoder used in the hierarchical embeddings module, 24,000 pairs of embeddings are selected at random from in each dataset in a 1:5 ratio of similar to dissimilar pairs. We chose this ratio since it is preferable to create overly specific log clusters than overly general log clusters, as this prevents the wrong template from being assigned to logs and combining multiple log clusters with the same template is comparatively easier. The code is implemented in PyTorch 2.3. During training, we use the Adam optimizer and chose an initial learning rate of 0.0005. We set the batch size to 2048 and train for 50 epochs. We also use a similarity threshold $T_c$ for log clustering of 0.9.

\section{EVALUATION RESULTS}
\begin{table*}[ht]
\centering
\caption{Accuracy of HELP compared to state-of-the-art baselines on public datasets.}
\label{tab:RQ1}
\resizebox{\textwidth}{!}{
\begin{NiceTabular}{c||c|cccccccccccccc|c}
\toprule
{Method} & {Metric} & {\small Apache} & {\small BGL} & {\small Hadoop} & {\small HDFS} & {\small HealthApp} & {\small HPC} & {\small Linux} & {\small Mac} & {\small OpenSSH} & {\small OpenStack} & {\small Proxifier} & {\small Spark} & {\small Thunderbird} & {\small Zookeeper} & {\small Average} \\

\midrule
\multicolumn{17}{c}{\textbf{Offline Log Parsers}} \\
\midrule

\multirow{4}{*}{AEL} & GA & \textbf{1} & 0.915 & 0.823 & \underline{0.999} & 0.725 & 0.748 & \underline{0.916} & 0.797 & 0.705 & 0.743 & \underline{0.974} & — & 0.786 & 0.996 & 0.860 \\
 & FGA & \textbf{1} & 0.587 & 0.117 & 0.764 & 0.008 & 0.201 & 0.806 & 0.793 & 0.689 & 0.682 & 0.667 & — & 0.116 & 0.788 & 0.560 \\
 & PA & 0.727 & 0.406 & 0.535 & 0.621 & 0.311 & 0.741 & 0.082 & 0.245 & 0.364 & 0.029 & 0.677 & — & 0.163 & 0.842 & 0.440 \\
 & FTA & 0.517 & 0.165 & 0.058 & 0.562 & 0.003 & 0.136 & 0.217 & 0.205 & 0.333 & 0.165 & 0.417 & — & 0.035 & 0.465 & 0.250 \\
\hline
\multirow{4}{*}{LogPPT} & GA & 0.786 & 0.311 & 0.533 & 0.694 & 0.839 & 0.782 & 0.200 & 0.536 & 0.278 & 0.534 & 0.51 & 0.450 & 0.416 & 0.973 & 0.560 \\
 & FGA & 0.484 & 0.519 & 0.570 & 0.385 & 0.863 & 0.812 & 0.686 & 0.459 & 0.115 & \underline{0.929} & \underline{0.750} & 0.383 & 0.334 & 0.903 & 0.590 \\
 & PA & 0.952 & 0.855 & 0.725 & 0.897 & \textbf{0.987} & \textbf{0.997} & 0.621 & 0.41 &\textbf{ 0.713} & 0.408 & \underline{0.993} & \underline{0.954} & 0.283 & \underline{0.843} & 0.760 \\
 & FTA & 0.352 & 0.509 & 0.462 & 0.308 & 0.787 & 0.762 & 0.399 & 0.275 & 0.147 & 0.788 & \underline{0.917} & 0.331 & 0.131 & 0.753 & 0.490 \\
\hline
\multirow{4}{*}{LILAC} & GA & \textbf{1} & 0.894 & 0.872 & \textbf{1} & \textbf{1} & \textbf{0.869} & \textbf{0.971} & \underline{0.876} & 0.690 & \textbf{1} & \textbf{1} & \textbf{1} & \underline{0.806} & \textbf{1} & \textbf{0.927} \\
 & FGA & \textbf{1} & \underline{0.859} & \textbf{0.962} & \textbf{0.968} & \underline{0.967} & \underline{0.907} & \textbf{0.931} & \underline{0.825} & \underline{0.838} & \textbf{1} & \textbf{1} & \textbf{0.901} & \underline{0.793} & \underline{0.967} & \textbf{0.924} \\
 & PA & \textbf{0.996} & \underline{0.958} & 0.832 & \textbf{0.999} & 0.687 & 0.705 & \textbf{0.765} & \underline{0.638} & \textbf{0.941} & \textbf{1} & \textbf{1} & \textbf{0.973} & \underline{0.559} & 0.687 & \textbf{0.842} \\
 & FTA & \textbf{0.862} & \underline{0.746} & \underline{0.779} & \textbf{0.946} & \textbf{0.868} & \underline{0.800} & \textbf{0.740} & \underline{0.553} & \textbf{0.865} & \textbf{0.979} & \textbf{1} & \underline{0.759} & \underline{0.572} & \underline{0.868} & \textbf{0.810} \\

\midrule
\multicolumn{17}{c}{\textbf{Online Log Parsers}} \\
\midrule
\multirow{4}{*}{LenMa} & GA & 0.993 & — & 0.796 & \underline{0.999} & — & 0.793 & 0.806 & 0.701 & \textbf{0.748} & 0.851 & 0.504 & — & — & 0.857 & 0.800\\
 & FGA & 0.900 & — & 0.059 & 0.831 & — & 0.035 & 0.616 & 0.107 & 0.800& 0.395 & 0.189 & — & — & 0.661 & 0.46 \\
 & PA & 0.031 & — & 0.052 & 0.137 & — & 0.642 & 0.035 & 0.121 & 0.142 & 0.019 & 0.495 & — & — & 0.683 & 0.24 \\
 & FTA & 0.267 & — & 0.009 & 0.404 & — & 0.015 & 0.17 & 0.021 & 0.200 & 0.06 & 0.151 & — & — & 0.263 & 0.16 \\
\hline
\multirow{4}{*}{Drain} & GA & \textbf{1} & \underline{0.919} & \underline{0.921} & \underline{0.999} & 0.862 & 0.793 & 0.686 & 0.761 & \underline{0.707} & 0.752 & 0.692 & \underline{0.888} & \textbf{0.831} & 0.994 & 0.840 \\
 & FGA & \textbf{1} & 0.624 & 0.785 & 0.935 & 0.01 & 0.309 & 0.778 & 0.229 & \textbf{0.872} & 0.007 & 0.206 & 0.861 & 0.237 & 0.904 & 0.550 \\
 & PA & 0.727 & 0.407 & 0.541 & 0.621 & 0.312 & 0.721 & 0.111 & 0.357 & 0.586 & 0.029 & 0.688 & 0.394 & 0.216 & \underline{0.843} & 0.470 \\
 & FTA & 0.517 & 0.193 & 0.384 & 0.609 & 0.004 & 0.152 & 0.259 & 0.069 & 0.487 & 0.002 & 0.176 & 0.412 & 0.071 & 0.614 & 0.280 \\
\hline
\multirow{4}{*}{UniParser} & GA & 0.287 & 0.549 & 0.718 & \textbf{1} & 0.447 & 0.794 & 0.263 & \textbf{0.886} & 0.495 & \textbf{1} & 0.51 & 0.853 & 0.440 & 0.996 & 0.660 \\
 & FGA & 0.256 & 0.641 & 0.490 & \underline{0.936} & 0.536 & 0.670 & 0.444 & 0.743 & 0.017 & \textbf{1} & 0.171 & 0.001 & 0.398 & 0.678 & 0.500 \\
 & PA & 0.456 & 0.949 & \underline{0.835} & \underline{0.948} & 0.814 & 0.949 & 0.171 & \textbf{0.685} & 0.474 & 0.516 & 0.634 & 0.775 & 0.331 & \textbf{0.989} & 0.680 \\
 & FTA & 0.093 & 0.223 & 0.319 & 0.574 & 0.303 & 0.371 & 0.255 & 0.281 & 0.011 & 0.292 & 0.229 & 0 & 0.176 & 0.500 & 0.260 \\


\midrule
\multicolumn{17}{c}{\textbf{Our proposed method}} \\
\midrule

\multirow{4}{*}{HELP} & GA & \underline{0.993} & \textbf{0.999} & \textbf{0.923} & \underline{0.999} & \underline{0.991} & \underline{0.854} & 0.774 & 0.817 & 0.625 & \underline{0.940} & 0.847 & 0.800 & 0.762 & \underline{0.999} & \underline{0.880} \\
 & FGA & \underline{0.931} & \textbf{0.961} & \underline{0.928} & 0.857 & \textbf{0.968} & \textbf{0.944} & \underline{0.923} & \textbf{0.869} & 0.722 & 0.896 & 0.727 & \underline{0.898} & \textbf{0.939} & \textbf{0.978} & \underline{0.896} \\
 & PA & \underline{0.969} & \textbf{0.996} & \textbf{0.908} & 0.690 & \underline{0.975} & \underline{0.990} & \underline{0.756} & 0.577 & 0.622 & \underline{0.946} & 0.939 & 0.934 & \textbf{0.644} & 0.669 & \underline{0.830} \\
 & FTA & \underline{0.754} & \textbf{0.918} & \textbf{0.792} & \underline{0.630} & \underline{0.842} & \textbf{0.908} & \underline{0.625} & \textbf{0.565} & \underline{0.686} & \underline{0.825} & 0.696 & \textbf{0.786} & \textbf{0.638} &\textbf{ 0.878} & \underline{0.753} \\

\bottomrule

\end{NiceTabular}%
}
\end{table*}%

\begin{table}[]
\small
\caption{Ablation study of HELP components.}
\label{tab:RQ2}
\begin{tabularx}{0.7\columnwidth}{l|c c}
    \toprule
    \textbf{Method} & \textbf{GA} & \textbf{FGA} \\
    \midrule
    HELP & 0.878 & 0.931 \\
    w/o LLM & $0.620_{(\downarrow 29.4\%)}$  & $0.891_{(\downarrow 4.35)}$ \\
    w/o prev. \& NN & $0.581_{(\downarrow 33.76\%)}$  &  $0.855_{(\downarrow 8.16\%)}$ \\
    w/o prev. \& WC & $0.535_{(\downarrow 39.05\%)}$ & $0.817_{(\downarrow 12.28\%)}$ \\
    w/o prev. \& IRM & $0.499_{(\downarrow 43.13\%)}$ & $0.804_{(\downarrow 13.71\%)}$ \\
    \bottomrule
\end{tabularx}
\end{table}

\subsection{RQ1: How effective is HELP?}
To determine the efficacy of HELP compared to other state-of-the-art log parsers, we evaluate all log parsers on the four metrics mentioned in ~\autoref{sec:met}. The results are shown in ~\autoref{tab:RQ1}, where the best result for each metric is marked in \textbf{bold}, while the second best result is \underline{underlined}. Following previous works \cite{jiang_large-scale_2023, khan_guidelines_2022}, if a parser cannot finish parsing in a reasonable time-frame (12 hours), its score is denoted as "—".

Our results show that HELP significantly outperforms all other online log parsers, and achieves the second best results overall, only performing slightly worse than the best offline parser LILAC. Additionally, HELP achieves consistently high accuracy across all 14 Loghub 2.0 datasets, achieving the best FTA on 14 of 14 datasets and the best FGA on 10 of 14 datasets compared to other online log parsers. This indicates HELP's robustness in log parsing data from a diverse set of log sources. 

In terms of group-related metrics (i.e. GA and FGA), Drain, the best online baseline achieves a GA of 0.840 and FGA of 0.550. HELP achieves 4.8\% and 62.9\% higher GA and FGA compared to Drain, with a GA of 0.880 and FGA of 0.896. Note that the FGA of Drain is much lower than HELP. This is because syntax-based parsers rely on hand crafted features that struggle to handle a large and diverse set of templates. On the other hand, HELP's use of semantic embeddings allows it to better catch edge cases and achieve superior performance on grouping accuracy. 

In terms of parsing-related metrics (i.e. PA and FTA), UniParser achieved the highest PA of 0.680 and the highest FTA of 0.260 among all online baselines. HELP achieves 22.0\% higher PA and 189.6\% higher FTA compared to UniParser, with a PA of 0.830 and a FTA of 0.753. This illustrates how leveraging the extensive pre-trained knowledge of universal LLMs is preferable to training a domain-specific model from scratch for log parsing. 

Compared with existing offline log parsers, HELP outperforms AEL and LogPPT by an average of 88.0\% and 43.0\% across all four metrics and achieves comparable performance to the existing state-of-the-art parser LILAC. Specifically, on datasets with a large number of templates like BGL (320), Hadoop (236), HealthApp (156), Linux (338), Mac (626), Spark (236), and Thunderbird (1,241), HELP outperforms LILAC by 4.0\% in FGA, 7.0\% in PA, and 3.0\% in FTA. This illustrates how fine-tuned semantic embeddings outperform LLM and token based methods on systems with large and diverse log templates. Because HELP is tuned to be more sensitive to create overly specific templates, HELP performs worse on datasets with fewer templates. This increased specificity helps reduce false negatives, which are more consequential in log parsing compared to false positives. This is because false negatives create silent errors that can cause system outages while false positives are shown as alerts that can be manually ignored by developers. Additionally, while LILAC is an offline parser that requires top-K frequent tokens for clustering and retrieval of similar templates during parsing, HELP achieves comparable performance while being an online parser where logs must be processed in real-time and the information available for parsing is incomplete. This highlights HELP's advantage in production environments, where real-time parsing is critical for downstream applications.
\begin{figure}[t!]
  \centering
  \includegraphics[width=\linewidth]{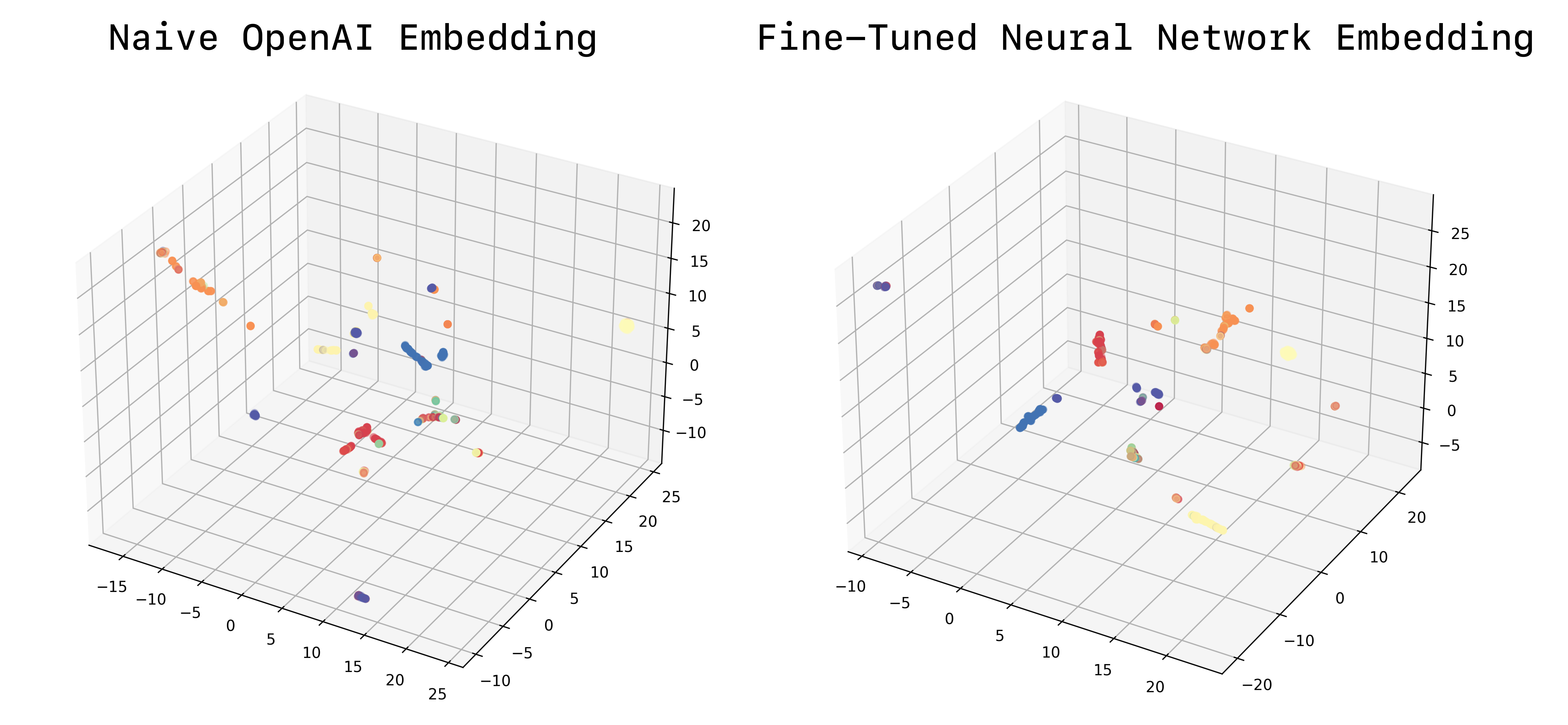}
  \caption{UMAP visualization of 1000 HealthApp log embeddings with 20 neighbors.}
  \Description{Comparison of 3D UMAP dimension reduction of 1000 HealthApp log embeddings for naive and fine-tuned log embeddings}
  \label{fig:UMap}
\end{figure}
\subsection{RQ2: How does each component contribute to HELP?}

\subsubsection{Ablations.} To determine the relative importance of each component of HELP in its overall performance, we conduct a series of ablation studies on six of the Loghub 2.0 datasets (BGL, Hadoop, HealthApp, Linux, Mac, and Thunderbird), which are chosen for their large template count ($>$ 100 templates). Specifically, we incrementally remove four components and compare their GA and FGA to the unmodified HELP. The four components we remove are 1) HELP w/o LLM: HELP removing the context-aware parsing module, 2) HELP w/o prev. \& NN: HELP removing the context-aware parsing module and replacing the fine-tuned neural network embeddings with the unmodified OpenAI embeddings 3) HELP w/o prev. \& WC: HELP removing all aformentioned parts and the embedding of the log word count data, 4) HELP w/o prev. \& IRM: HELP removing all previously mentioned components and the iterative rebalancing module. 

The evaluation results are shown in ~\autoref{tab:RQ2}. The results make the importance of each design choice of HELP clear. For instance, removing the context-aware parsing module significantly impacts GA while having a smaller effect on FGA. This difference arises because the hierarchical embedding module tends to create overly specific groups, which has a greater impact on GA, a log-level metric, compared to FGA, a template-level metric. The parsing module mitigates this issue by merging these overly specific groups by assigning them to the same template, improving overall accuracy. Our hypotheses regarding the importance of domain-specific log information, the limitations of semantic models, and the detrimental effects of log drift in online algorithms are further validated by the noticeable performance declines when the fine-tuned neural network embeddings, word count, and iterative rebalancing module are removed.

\subsubsection{Visualization.} One of the main contributions of our paper is demonstrating that embedding models provide a cost-effective and efficient alternative to sequential LLM querying. To validate the capability of embedding models in accurately creating log groupings, and to demonstrate the efficacy of fine-tuned neural network embeddings for this task, we visualize 1000 embeddings from the HealthApp dataset using Unifold Manifold Approximation and Projection (UMAP) \cite{mcinnes_2020_umap}, a dimension reduction technique. The results are shown in ~\autoref{fig:UMap}. The left panel shows the UMAP visualization of naive OpenAI embeddings, while the right panel displays the fine-tuned neural network embeddings. Although both are able to capture template groupings, the more densely grouped clusters and the greater separation between clusters of different templates from the fine-tuned embeddings highlights the effectiveness of fine-tuning in improving log groupings.

\begin{figure}[t!]
  \centering
  \includegraphics[width=\linewidth]{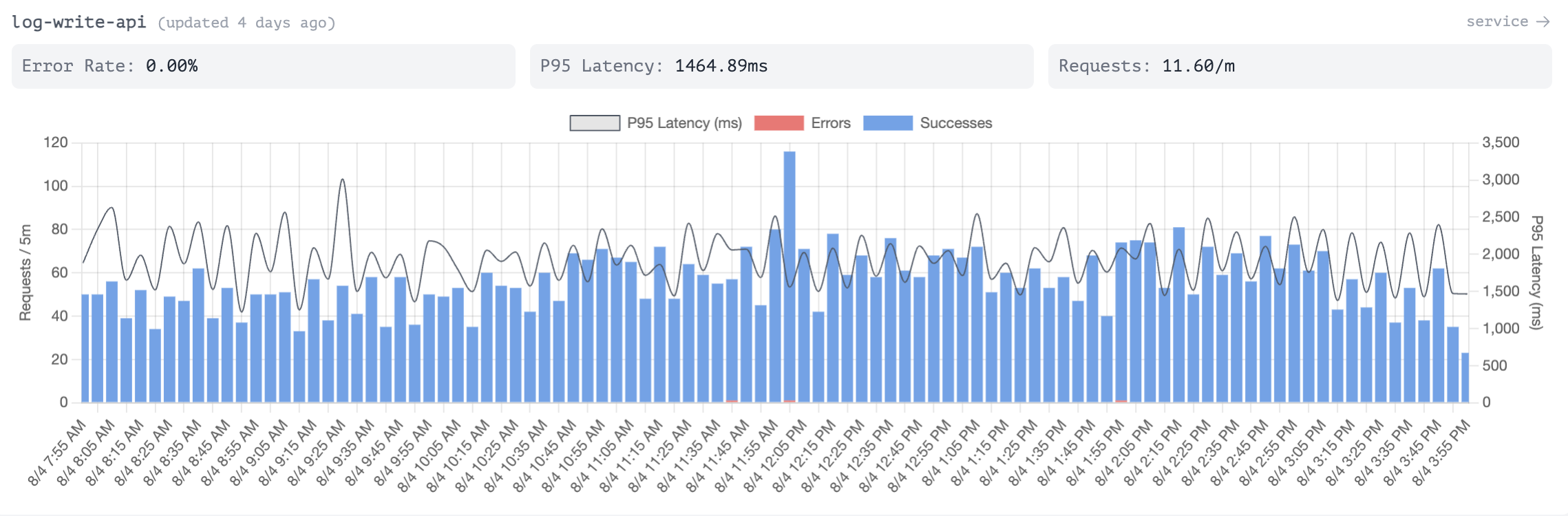}
  \caption{Latency statistics of Iudex log write service.}
  \Description{Latency and request information}
  \label{fig:Latency}
\end{figure}
\subsection{RQ3: How practical is HELP in a production environment?}

HELP has been successfully deployed into Iudex's production observability platform as an online algorithm. Because multiple services emit logs concurrently, HELP is implemented as a batched algorithm, where thousands of logs can be embedded and grouped in parallel. Internal testing has found that HELP's iterative rebalancing module allows HELP to achieve little to no performance degredation as a batched algorithm, with an average decrease of 1-2\% in grouping accuracy while providing significantly lower latency. ~\autoref{fig:Latency} shows the end-to-end P95 latency from log emission to log storage and grouping over an eight hour time-frame. We achieve an average P95 latency of approximately 1.5 seconds, making HELP efficient and effective for real-time log analysis tasks.

\section{DISCUSSION}
\subsection{Practicality of LILAC}
HELP is designed to be a universal model for log parsing. Although HELP requires an initial offline pre-training step on labeled log data, after pre-training, HELP can be applied to new services for online inference without any additional supervised data or training. We provide the neural network model weights for HELP trained on all 14 datasets for other organizations interested in utilizing HELP's fine-tuned embeddings, bypassing the need for their own pre-training. As we have shown by successfully deploying HELP into Iudex, HELP's use of cost effective text-embeddings makes real-time log parsing practical for production deployment. 
\subsection{Threats to Validity}
\textbf{Data Leakage.} Because LLMs are trained extensively on a vast corpora of text data, there exists a possibility of data leakage where the embedding or  language models have previously seen the open source log datasets. This could result in the model memorizing ground-truth templates during inference. However, our superior performance against methods such as LogPPT which solely use LLMs, our ablation studies which remove individual components of HELP, and the robust performance of HELP in private production data indicate a low probability of rote memorization.

\textbf{Randomness.} Randomness can affect the performance of HELP and other baseline methods. To combat this, HELP minimizes the randomness of its context-aware parsing module by setting the temperature of the model to 0, ensuring consistent outputs for the same input text. HELP also run each experiment three times and averages the results to reduce variance.
\section{CONCLUSION}
Log parsing is the foundation for log analysis tasks, whereby semi-structured log data is transformed into structured log templates. In this paper, we propose a novel hierarchical embeddings-based log parser, HELP. HELP is the first online semantic-based parser to leverage LLMs for effective and efficient log parsing. HELP achieves this via its hierarchical embeddings module, which fine-tunes a text embedding model to cluster logs before parsing, reducing inference costs by multiple orders of magnitude. To combat log drift which can create false positive alerts and to enable deployment in parallel, HELP implements an iterative rebalancing module to periodically update existing log groupings. We evaluate HELP extensively on 14 large scale public datasets, where HELP significantly outperformed other online parsers on grouping and parsing accuracy. Finally, HELP is deployed into Iudex's production observability platform, demonstrating HELP's practicality in high-throughout real-world log parsing.



\bibliographystyle{ACM-Reference-Format}
\bibliography{RefTex}
\end{document}